\documentclass[sigconf]{acmart}
\renewcommand\footnotetextcopyrightpermission[1]{} 
\usepackage{tikz}
\usepackage{amsmath}
\usepackage{tabularx}
\newcolumntype{Y}{>{\centering\arraybackslash}X}

\newcolumntype{Z}{>{\hsize=1.2\hsize}X}
\newcolumntype{Q}{>{\hsize=.8\hsize}X}
\newcolumntype{V}{>{\hsize=.15\hsize}X}

\usepackage{caption}
\usepackage{subcaption}
\usepackage{stfloats}
\usepackage{multirow}
\usepackage{multicol}
\usepackage{makecell}
\usepackage{calc}  
\usepackage{enumitem} 
\usepackage{nicematrix,tikz}
\usepackage{xurl}

\usepackage{filecontents}

\AtBeginDocument{%
  \providecommand\BibTeX{{%
    \normalfont B\kern-0.5em{\scshape i\kern-0.25em b}\kern-0.8em\TeX}}}

\setcopyright{acmcopyright}
\copyrightyear{2018}
\acmYear{2018}
\acmDOI{XXXXXXX.XXXXXXX}

\acmConference[Conference acronym 'XX]{Make sure to enter the correct
  conference title from your rights confirmation emai}{June 03--05,
  2018}{Woodstock, NY}
\acmPrice{15.00}
\acmISBN{978-1-4503-XXXX-X/18/06}

\begin{document}

\title{Folk Models of Misinformation on Social Media}


\author{Filipo Sharevski}
\affiliation{%
  \institution{DePaul University}
  \streetaddress{243 S Wabash Ave}
  \city{Chicago}
  \state{IL}
  \postcode{60604}
    \country{United States}
}
\email{fsharevs@cdm.depaul.edu}

\author{Amy Devine}
\affiliation{%
  \institution{DePaul University}
  \streetaddress{243 S Wabash Ave}
  \city{Chicago}
  \state{IL}
  \postcode{60604}
    \country{United States}
}
\email{adevine@depaul.edu}

\author{Emma Pieroni}
\affiliation{%
  \institution{DePaul University}
  \streetaddress{243 S Wabash Ave}
  \city{Chicago}
  \state{IL}
  \postcode{60604}
    \country{United States}
}
\email{epieroni@depaul.edu}

\author{Peter Jachim}
\affiliation{%
  \institution{DePaul University}
  \streetaddress{243 S Wabash Ave}
  \city{Chicago}
  \state{IL}
  \postcode{60604}
    \country{United States}
}
\email{pjachim@depaul.edu}

\renewcommand{\shortauthors}{Authors}

\begin{abstract}
In this paper we investigate what \textit{folk models of misinformation} exist through semi-structured interviews with a sample of 235 social media users. Work on social media misinformation  does not investigate how ordinary users---the target of misinformation---deal with it; rather, the focus is mostly on the anxiety, tensions, or divisions misinformation creates. Studying the aspects of creation, diffusion and amplification also overlooks how misinformation is internalized by users on social media and thus is quick to prescribe ``innoculation'' strategies for the presumed lack of immunity to misinformation. How users grapple with social media content to develop ``natural immunity'' as a precursor to misinformation resilience remains an open question. We have identified at least five \textit{folk models} that conceptualize misinformation as either: \textit{political (counter)argumentation}, \textit{out-of-context narratives}, \textit{inherently fallacious information}, \textit{external propaganda}, or simply \textit{entertainment}. We use the rich conceptualizations embodied in these folk models to uncover how social media users minimize adverse reactions to misinformation encounters in their everyday lives. 
\end{abstract}




\maketitle

\section{Introduction}
The proverbial ``fool me once, shame on you'' placed the shame on foreign actors as the culprit in spreading misinformation on social media as an act of fooling the electorate in US and UK in the mid 2010s \cite{Benkler, Tumber}. Researchers, governments, the press, think tanks, NGOs all blamed these foreign actors---bots, trolls, or clickbait farms \cite{ZannettouS}---for deceitfully substituting the truth with relativism as a pillar of these liberal democracies \cite{Farkas, Waisbord}. Then, a response to this conundrum followed. Twitter produced public datasets from their information operations accounts and content \cite{Twitter-IO}, Facebook got under scrutiny for enabling and amplifying misinformation \cite{Berghel}, legislators explored laws and regulations \cite{Goldberg}, and academia focused on structural containment of misinformation \cite{Im, Linden, Jachim}. All of these efforts were driven towards \textit{detecting}, \textit{removing} or \textit{flagging} misinformation, as ``fool me twice, shame on me'' was not going to happen; researchers blamed  misinformation for truth decay \cite{Rich}, societal anxiety \cite{Verma}, and heightened interpersonal polarisation \cite{Stewart}. As such, misinformation and the ``misinformers'' became a direct threat to the liberal democratic order \cite{Wigell} (later in time, to the public health too \cite{Cinelli}).   

And the one being fooled for the second time was not just the social order and the establishment, but also the individual constituents, i.e. the everyday users of social media. 
Researchers deemed social media users
Past research has deemed social media users too biased \cite{Flynn}, lazy, deluded, ignorant \cite{Pennycook1, Bronstein}, or simply illiterate \cite{Buhler} to be able to deal with misinformation so we were to be taught how to \textit{recognize} it and ultimately \textit{reject} it \cite{Grace, Gimpel}. For us, the people of these liberal democracies, misinformation rejection is the only possible outcome for dealing with this ``information-based threat,'' from disinformation and fake news to rumors, hoaxes, and conspiracies theories \cite{Wu}. We are to follow training protocols \cite{Soetekouw}, play serious games \cite{Basol}, follow visual dashboards \cite{Lee}, and join virtual reality escape rooms \cite{Paraschivoiu} to make sure the shame does not fall on us, come what may next on the misinformation front. 

The good news about the bad news is that we have help in determining 
\textit{how accurate} our mental process of dealing with misinformation. The not so good news about the bad news is that no work so far has studied what mental processing models exist in dealing with misinformation from a folk perspective in the first place, and how well these models \textit{serve a user} when grappling with misinformation. The top-down approach of misinformation containment might be good in theory, but it has difficulties to work well in practice. Users' mental processing largely fails to consider information accuracy corrections \cite{Swire-Thompson}, ignores misinformation warnings \cite{SPAM, Clayton}, and doesn't investigate misinformation on social media \cite{Geeng}. For these reasons, our project challenges the top-down approach of helping people process misinformation with an attempt to better how this process actually looks in the first place. 

We introduce the paradigm of \textit{misinformation folk models on social media} as an effort to explore how people respond to misinformation through contextualization, interpretations, and various forms of actions. The inductive articulation of the folk models of misinformation, for us, also serve a bigger purpose in approaching misinformation as an adverse situation, an information-based threat of ``fooling'' oneself. The conscious, albeit subjective, acknowledgement of an adverse situation forces people to decide how to response to misinformation, learn from experiences, and develop strategies for dealing with emerging (mis)information. The outcome, as much as it descriptively uncovers the current misinformation folklore, is an essential starting point in learning how people are predisposed, in not already stated, to build \textit{resilience to misinformation}. The later goal of our project is to address the already underway top-down mass ``inocculation'' strategizing for misinformation (interestingly, the pandemic misinformation made it hard not to introduce public health metaphors in response to the ``infodemic'') \cite{Lewandowsky, Jeon, Maertens}. 

Through a study with 235 social media users in United States, we found out that misinformation is conceptualized in several distinct ways based on how the information on social media relates to facts known to the public. Looking top-down, one would expect that most, if not all, of the participants in our study will possess what we named the \textit{inherently fallacious information} folk model:  misinformation is any information \textit{unfaithful} to to known facts, regardless of contexts or intentions (see \cite{VoxPop} for the discussion on the ``faithfulness to known facts'' qualification). However, the most prevalent folk model was the \textit{political counter(argumentation)}:  misinformation was conceptualized as any information with faithfulness to \textit{selective} facts relative to political and ideological contexts, created and disseminated with a political agenda-setting or argument-winning intentionality.

The second most prevalent folk model was the one where misinformation referred to all \textit{out-of-context narratives} with a \textit{questionable} faithfulness to known facts due to selection of improbable alternative contexts, created and disseminated with speculative intentions. In the political counter(argumentation) and the out-of-context narratives folk models, facts \textit{do appear} as part of the misinformation, but the appearance is predicated to a certain extent by self-promoting goals of other users on social media, instead of misinformation being only the \textit{disappearance} of facts caused by the foreign actors. The foreign actors enter the picture with the \textit{external propaganda} folk model, where misinformation is any information with a \textit{fluctuating} faithfulness to known facts relative to shifting contexts, created and disseminated with a propagandistic intentions. We found that the users did not just expect to be ``fooled'' but also developed an \textit{entertainment} folk model where misinformation is any information with a \textit{tangential} faithfulness to known facts relative to humorous, sarcastic, or memetic contexts, usually created and disseminated with entertaining intentions.

As to who the ``misinformers'' are, the participants in our sample did not just point at foreign actors; they identified the following originators: users from the ``other'' side, die-hard speculator users, mass (social) media companies themselves, as well as the domestic establishment. The purpose of misinformation was not necessarily to ``fool'' a user on social media, but to: serve as political ammunition, stir the pot, increase profit, and entertain. When users in our sample were exposed to misinformation on social media, they employed variants of the dual model of information assessment proposed in \cite{Metzger} that incorporates both \textit{analytical} processing (e.g. reference to scientific data, fact checking) and \textit{heuristic} processing (e.g. gut feelings, linguistic formatting). Rejection of misinformation was already in place, not in a form of top-down prescribed dismissal in the name of the truth, but with far more versed tactics and actions. We identified 21 tactics within the five folk models in our study that users employ when responding to misinformation, from direct challenging to user-initiated policing of the misinformers. 

In a rather bottom-up fashion, the five folk models helped the users to begin developing a ``natural immunity'' through ``self-innoculation'' against misinformation on social media. A great deal of the participants, as analyzed further in the remainder of the paper, are already able to  ``bounce back'' with participation in the social media discourse i.e. their exposure to misinformation had no adverse self-reported reaction to them. Before we present the results in Sections 3-8, we examine misinformation as a wicked problem that resists exclusively a top-down approach in resolving it in Section 2. Using the results, we discuss possible alternative approaches of tackling the misinformation ``wickedness'' in Section 9 and conclude the paper in Section 10.


\section{Is Misinformation A Wicked Problem?}
A ``wicked problem'' is a problem which is ill--formulated, where information is confusing, with many clients and decision makers with conflicting values, and where ramifications in the whole system are thoroughly confusing. The adjective ``wicked'' is supposed to describe the mischievous and even evil quality of these problems, where proposed ``solutions'' often turn out to be worse than the symptoms. \cite{Rittel, Churchman}. This definition fits, at least at a first glance, when \textit{misinformation} is treated as a social problem. Treating misinformation as a social problem is not hard to justify, as information around social events, elections, public health, and even wars is scrutinized for accuracy and source credibility \cite{Shu}, truthfulness \cite{Zhang}, and regularly fact-checked \cite{Tambuscio} (at least these became widespread practices after the 2016 US presidential elections and UK Brexit campaign). 


The resonance with ``wickedness'' intensifies further when attempts are made to formulate ``misinformation.'' Is misinformation information that is: (i) fake \cite{Wu}; (ii) false \cite{ZannettouS}; (iii) unfaithful to known facts \cite{VoxPop}; or (iv) deceptive fabrication \cite{Starbird}? One is tempted to say this is all the same, but a close look reveals subtle yet important differences that complicate the treatise of misinformation \cite{Molina}. If misinformation is fake information, then it requires a case-by-case determination of the likelihood that the fabrication is misconstrued as factual because ``hardly any piece of `fake news' is entirely false, and hardly any piece of real news is flawless'' \cite{Potthast}. If misinformation is outright false or ``untrue,'' it in turn requires a grounding of a universal truth, which can be a difficult endeavor dependent on collective consensus \cite{Southwell}. The difficulty arises from the radically subjective interpretation of the ``truth'' and entails considering the degree of faithfulness of a piece of information to known facts at a given point in time because facts could change, be refuted, or reformulated \cite{VoxPop}. But one could argue that faithfulness is also an ill-defined concept as it requires plurality of facts, leading to further confusions of epistemic relativism \cite{Higgins}. 

To add to the confusion and complication, misinformation also requires consideration of the intent to create and disseminate any content for the purpose of deceit, manipulation of public opinion, or substantiating hard-line beliefs. Misinformation, according to the definition in \cite{Allcott}, encompasses information that could mislead readers and is intentionally and verifiably false (one would say this is ``disinformation,'' an important distinction we also acknowledge as part of the typology facet of the overall complexity \cite{Wu, Starbird, ZannettouS}). The intentionality is therefore a discriminative token that deems satire as misinformation as it is intentionally fabricated content (e.g. Onion news) but excludes unintentional misreporting, rumors, or conspiracy theories that are not entirely false. Satire proponents could disagree and counter-argue that satire could include some verifiable facts and the intention is to entertain, and not to deceive users \cite{Rubin}. A disagreement around intentionality exemptions, if not for the rumors and conspiracy theories, exists when mainstream online news media unintentionally misreport the news. Despite outlets having the ability to issue an errata or corrections to misreporting, evidence suggests they nonetheless routinely disseminate verifiable false information surrounding elections, public health issues, and foreign affairs \cite{Benkler}. In these instances, the editorial pressure to report with speed, while engaging audiences in the process, is a `solution' worse than misinformation `symptoms.' 

Solutions for misinformation symptoms do not get better even when the speed, audience engagement, and the power of the crowds are brokered on social media platforms. To begin with, social media platforms themselves hold conflicting values regarding misinformation as the mainstream ones consider misinformation a violation of the end user agreement for platform participation while the alternative ones treat misinformation as another type of information in the ``marketplace of ideas'' \cite{gettr-paper}. Twitter, for example, defines misinformation ``as claims that have been confirmed to be false by external, subject-matter experts or include information that is shared in a deceptive or confusing manner'' and pledges to ``remove, label, or prebunk'' such claims \cite{twitter-misinfo}. Meta even touches upon the intentionality aspect, stating that ``tackling misinformation actually requires cracking down on several challenges including fake accounts, deceptive behavior, and misleading and harmful content'' \cite{meta-misinfo}. But the mainstream solutions of administrative content/account moderation to curb misinformation are seen as overtly intrusive by the alternative platforms and they instead take a hands-off approach. Gettr, for example, brandishes an image of a ``marketplace of ideas'' where users should expect the truth to emerge without any administrative censorship \cite{gettr}. Parler similarly offer(ed) users a venue to ``express openly, without fear of being administratively deplatformed for their views'' \cite{Parler}, Gab ``champions free flow of information online'' \cite{Gab}, and 4chan allows ``anyone to post comments and share images'' ~\cite{4chan}. 

One would reckon that doing at least something (moderation, removal, censorship, prebunking, debunking) is better than doing nothing when misinformation becomes ingrained across the social media space. Eventually, these `anti-misinformation' strategies will yield the desired outcome, but for now, they seem to do the opposite and exacerbate the problem further. Misinformation labels and prebunking are found on numerous occasions to make users believe misinformation on mainstream platform more, not less, causing a so-called ``backfiring effect'' \cite{flag-paper, Clayton, Nyhan, Ecker, Thorson}. And the problem is not just whether to apply moderation or not, but also how frequent and in what context should misinformation be dealt with. Too frequent moderation paired with constant pre/debunking is also found to create an ``illusory truth effect'' \cite{Pennycook1}, while the absence of the moderation in some scenarios creates an `implied truth effect'' and lead users to deem any misinformation content they encounter as credible and accurate \cite{Pennycook}.

The attempted control of the `supply' of misinformation through editorial checks or content/account moderation needs also to consider the `demand' for misinformation from the users. The assumption that the users want the truth, the whole truth, and nothing but the truth might not hold as the misinformation is entrenched in the partisan agenda and users battle with alternative narratives on almost every issue \cite{Wilson}. Studies emphasize that the proclivity for alternative narratives is  most salient among the right-leaning users \cite{Bhatt}, but new user evidence from the alt-platform Gettr reveals that misinformation is equally attractive for left-leaning or moderate users \cite{gettr-paper}. While right-leaning users claim disenchantment with `oppressive' free speech infringement by mainstream social media, left-leaning users are drawn to misinformation to `keep abreast with the latest argumentation of the right online.' As counter-intuitive as it seems, misinformation is an invaluable tool for driving partisan engagement and many `clients' might be worse-off if misinformation suddenly vanished.

To come full circle, what might seem like `strategic information operations' targeting passive users at the beginning, could be a galvanizing force that attracts more active misinformation consumers/producers and more participation every time a social issue arises. The calculated and often too ambiguous institutional response to misinformation does little to rectify this `wickedness' as it frames misinformation as an exogenous threat and largely ignores both its regulatory responsibility and the domestic supply/demand chain of misinformation \cite{Wood}. Aside from `myth busters' around COVID-19 and election ballot processing, online users could seek little guidance from the institutions in discerning misinformation. Discerning misinformation is, according to current evidence \cite{Kirchner, Soetekouw, Gimpel}, not something users are endowed with, and experts for now only provide strategies for \textit{inoculation} against misinformation \cite{Lewandowsky, Jeon} in addition to the automated detection \cite{Pelrine, Jachim, Vargas}.

The combination of `inoculation' and automated detection could help counter the `wickedness' and is definitely worth pursuing. But this approach is limited in that users must supply their own context and interpretations of every new misinformation content or alternative narrative. In such conditions, investigating what \textit{folk models of misinformation} are employed when dealing with misinformation is warranted to understand perils to people against which at least the `inoculation' is supposed to work \cite{Peck}. What do \textit{people} actually think of misinformation and how do they respond to it has never been the main objective of inquiry as folk models are always associated with often ``inaccurate representation of the real world and thus lead to presumably erroneous decision-making (or the assumption that people are bad in dealing with misinformation) ~\cite{Wash-folk, Torres}.  

But whether the folk models are correct or not is immaterial to countering the ``wickedness'' of misinformation as both the `inoculation' strategies and the automated detection should be designed to work well with the folk models of misinformation actually employed by users in the first place. This line of argumentation is not strange when dealing with technology in general and cybersecurity in particular \cite{Camp}. Understanding the folk models of computing, networking as well as malware, passwords, privacy and cybercrime helped bring cybersecurity technologies and `inoculation' against hacking to markedly improve synergy with users \cite{Ur, Volkamer, Kang, Marki}. Naturally, one would expect that an attempt of doing so when dealing with misinformation could be beneficial too.   

\section{Folk Models of Misinformation on Social Media} 
\subsection{Research Questions}

Developing an understanding of what mental models people actually possess is not to generalize to a population, but rather to explore a phenomenon in depth \cite{Wash-folk}. In that, the goal of our study is not to overcome the ``wickedness'' in a single shot, but take a bottom-up approach to address it. This approach is counter to the dominant, top-down approach approach in the ``war of misinformation'' where generalizations of users' inadequacy, social media complacency, and ``us vs them'' mentality are prevalent. To this point, we focused on seven research questions respective to how users think and act upon misinformation in their daily engagement on social media:  

\begin{enumerate}
\itemsep 0.5em
    \item \textit{Definition}: How do social media users conceptualize misinformation (i.e. what is misinformation according to them)? 
    \item \textit{Origins}: Who do social media users believe creates and benefits misinformation? 
    \item \textit{Purpose}: How do social media users interpret the purpose of misinformation?
    \item \textit{Assessment}: How do social media users decide whether given content is or is not misinformation?
    \item \textit{Tactical Response}: What tactics do social media users employ when responding to misinformation?
     \item \textit{Resilience}: How might social media users respond to emergent misinformation?
\end{enumerate}

\subsection{Sample}
Our study was approved by the Institutional Review Board (IRB) of our institution before we conducted the semi-structured interviews (the questionnaire is provided in the Appendix). We sampled a population that was 18 years or above old, from United States, that regularly uses social media, and has encountered misinformation on social media platforms. We used Amazon Mechanical Turk to recruit our participants. After we removed low quality responses, we ended with a sample of total of 235 participants. The semi-structured interviews were anonymous, allowed users to skip any question they were uncomfortable answering, took around 40 minutes to complete it, and participants were offered a compensation rate of \$9.2 per hour.

The distribution of participants per their self-reported gender identity was 102 (43.4\%) were female, 117 (49.78\%) male, and 16 (6.82\%) preferred not to say. Age-wise, 32 (12.76\%) were in the [18-30] bracket, 100 (42.55\%) in the [31-40] bracket, 60 (25.53\%) in the [41-50] bracket, 28 (11.91\%) in the [51-60] bracket, and 15 (6.38\%) were 61+ old. The distribution of the political leanings within the sample was: apolitical 10 (4.25\%), left-leaning 115 (48.93\%), moderate 61 (25.95\%), and 49 right-leaning (20.85\%). Facebook was the platform where most of the participants (156) encountered misinformation, closely followed by Twitter (131 participants), Reddit (54 participants), 4chan (17 participants), and Gab, Parler, Gettr (combined 6 participants).

\subsection{Method and Analysis}
Participants were first provided an all-encompassing, academic characterization of misinformation synthesized from \cite{}. Asked to consider this characterization, participants were asked to describe their experience with encountering misinformation of social media platforms. This included open ended questions on the name of the platform, examples of misinformation, as well as participants' initial response. Next, we asked participants to provide their opinions on where information comes from, what the purpose of misinformation serves on social media, and who creates and benefits from it. Participants were then asked to further elaborate how do they suspect a certain social media post is misinformation, and what tactics they employ when dealing with misinformation. 

The qualitative responses were coded and categorized in respect: a) salient features that form a folk model; b) origins of misinformation; c) misinformation purpose; d) misinformation assessment, and e) tactical response to misinformation. Two independent researchers analyzed the raw responses received, achieving a strong level of inter-coder agreement---Cohen's $\kappa = .86$). We performed a basic exploratory analysis of each of the aforementioned aspects to uncover how social media users deal with misinformation on social media and develop resilience (``natural immunity'') to it. In reporting the results, we utilized as much as possible verbatim quotation of participants' answers, emphasized in ``\textit{italics}'' and with a reference to the participant as either \textbf{PX} or [\textbf{PX}], where \textbf{P} denotes ``\textbf{participant} and ``\textbf{X} denotes the \textbf{number} of the participant in the sample, e.g. \textbf{P32} indicates to \textbf{participant 32} (ordered by the time of participation).  

\section{What is Misinformation?}

\subsection{Political (Counter)Arguments}
The predominant conceptualization of misinformation within our sample refers to any information that has faithfulness to \textit{selective} facts relative to political and ideological contexts, created and disseminated with agenda-setting or argument-winning intentionality. This folk model of misinformation, in contrast to the `strategic information operations' \cite{Starbird}, moves the social media users from passive consumers to active partakers that shape misinformation with competing interpretations from the political information environment. In this sense, social media users are drifting away from regurgitating `state-sponsored troll farms' misinformation \cite{Caulfield} and instead focus on the rhetoric of political personas, commentators, think-tanks, and vocal supporters. In the effort to ``\textit{avoid admitting political defeat at any cost}'' as participant [\textbf{P65}] put it, the intentionality of the political partakers shifts from massive public opinion manipulation to ``\textit{hammering faulty logic into selection of facts as long as that discredits the `other' side}'' and ultimately wins an argument [\textbf{P16}].   

Participants' concept of misinformation as ``\textit{political statements that exaggerate the party agenda}'' [\textbf{P137}] is driven towards misleading suggestions about a preferred political stance, rather than towards promulgation of absolute falsehoods and inaccuracies \cite{Wu}. Participants acknowledge that ``\textit{falsehoods, inaccuracies, and unsourced claims could appear}'' [\textbf{P33}] in the the political (counter) argumentation, but they are neither exclusive nor always ``\textit{outlandish}'' [\textbf{116}], as was the case in past misinformation campaigns targeting political issues \cite{ZannettouS}. Facts do play a role in the folk concept of political (counter)argumentation as misinformation as it is ``\textit{taking little soundbite facts completely out of context and spreading them like wildfire}'' [\textbf{P213}]. People, thus, do not necessarily think of `fatefulness' to known facts \cite{VoxPop}, instead favoring ``\textit{selective facts}'' [\textbf{P61}] when other users ``\textit{piece together their opinion and put it on social media}'' [\textbf{P31}]. The responses, in general, uncover a trend where the proverbial `entitlement to one's own opinion \textit{but} not one's own facts' seems to become an `entitlement to one's own opinion \textit{and} one's own facts' as a concept of misinformation on social media, fostered by the need for ``\textit{reinforcement of ones political beliefs}'' [\textbf{P59, P214}] \cite{Lima}.    

\subsection{Out-of-context Narratives}
In this folk model, misinformation is any information that has \textit{questionable} faithfulness to known facts due to selection of improbable alternative contexts, created and disseminated with speculative intentionality. `Alternative  contexts' refers to an explanation of or narrative around an event or issue at state that runs counter to the mainstream  context and attempts to displace any factual reporting or development around the said event or issue \cite{Starbird2, Albright}. The alternative explanation and narratives in the past were ridden with fabricated information and encompassed wild political rumors, far-fetched conspiracy theories, and downright insinuations on issues readily verifiable with facts and simple methods. Social media users in our study do confirm that alternative narratives of such format do still appear on social media, but are not as prevalent as ones where the fact verification, plausibility assessment, and eliminating competing explanations is not straightforward. 

When thinking of the misinformation, these participants conceptualized it as ``\textit{facts with missing, incomplete, or made-up context}'' [\textbf{P201}], ``\textit{cherry picking events presented out of context in order to support a biased argument}'' [\textbf{P29}], or ``\textit{political campaign messages that include quotes and claims taken out of context}'' [\textbf{P110}]. Scientific facts or results were also conceptualized as `alternative narratives' as misinformation was also thought of as ``\textit{bad research}'' [\textbf{P191}], ``\textit{misconstrued study results interpreted in limited context}'' [\textbf{P189}], or ``\textit{conclusions based on incomplete evidence}'' [\textbf{P87}]. The out-of-context narratives could involve political events such as elections, but we classified this line of reasoning as a separate folk model because our participants mostly pointed to misinformation centered around topics such as health  (e.g. vaccines, COVID-19, medical/medicinal side-effects), foreign conflicts (e.g. the war in Ukraine) or other issues (e.g. Black Lives Matter, abortion rights, cryptocurrencies, celebrities, influencers).  

\subsection{Inherently Fallacious Information}
The concept of misinformation as exclusively ``\textit{information that is inherently fallacious}'' [\textbf{P64}] was perhaps not as dominant as one would expect. The two conceptualizations above acknowledge that misinformation does in many respects involve inaccuracies and falsehoods, but they are driven towards ``\textit{wild speculation}'' [\textbf{P22}] rather than fabrication for fabrication sake. Many identified misinformation as ``\textit{hoaxes that circulate on social media}'' [\textbf{P70}] that include ``\textit{fear mongering information}'' \textbf{[P120]}. Misinformation as ``\textit{lies}'' for political purpose [\textbf{P3}] or ``\textit{trolling of innocent users of the platform}'' [\textbf{P1658}] was also mentioned by our participants in reference to ``\textit{factually or scientifically incorrect}'' information [\textbf{P175}]. The inherent fallaciousness for this folk model conceptualizes misinformation as any information unfaithful to known facts, regardless of contexts or intentionality \cite{VoxPop}. 

\subsection{External Propaganda}
Unlike the previous folk model, the conceptualization of misinformation as ``\textit{external propaganda}'' [\textbf{P82}] refers to information that ``\textit{fluctuates}'' [\textbf{P153}] its faithfulness to known facts relative to shifting contexts or perceived profit-making/division-creating intentions. It could be said that the same conceptualization applies to the political (counter)argumentation or the out-of-context narratives, but we consider this one as a separate folk model that has more of an explicit ``\textit{propaganda}'' [\textbf{P210}] flavor to information. In the mind of our participants, the external propaganda misinformation works towards realizing agendas of ``\textit{mainstream (social) media companies},'' [\textbf{P4}], ``\textit{dark money and government intelligence agencies}'' [\textbf{P151}], or ``\textit{nation-states}'' [\textbf{P21}]. 

A distinguishing feature for this folk model is that the decision of whether information is or is it not misinformation is not explicitly in relationships to verifiable facts, but decided based on the actor(s) that produce and disseminate this information. Within this folk model, a piece of information \textit{entirely based on factual evidence} might be considered outside interference and automatically rejected if it appears to be from ``\textit{organized propaganda campaigns}'' [\textbf{P70}]. Participants in our sample assigned these campaigns to Twitter, Facebook, and mainstream news media corporations, deeming any information on these platforms or outlets to be ``\textit{leftist lies}'' [\textbf{P221}]. In equal measure, the outside interference coming from the ``conservative echo chambers of Fox, Newsmax, OANN, PraegerU, 4chan, Gab'' [\textbf{P103}] was considered responsible for spreading ``right-wing nonsense'' [\textbf{P226}].  

The ``\textit{Russian bots}'' [\textbf{P75}] are still mentioned in discussions of misinformation on social media, but our participants broadened the nation-state list to include ``China and Iran'' too [\textbf{P61}]. In this folk model, social media users' involvement is elevated from passive participation to active engagement of the nation-states' agenda to ``cause discord and chaos in public life'' [\textbf{P64}]. An interesting aspect of this folk model is that it outlines a \textit{chain of misinformation} where: a) ``\textit{the fake accounts controlled by the state actors put initial rumors and fabricated facts}'' on social media \textbf{[P95]}; b) these are picked up and amplified by ``\textit{demagogue figureheads that glom onto misinformation that suits their needs}'' [\textbf{P61}]; and c) kept alive by ``\textbf{ignorant users}'' [\textbf{P120}] that appropriated the misinformation as the ``\textit{preferred truth}'' [\textbf{P106}] fitting their biases or lack of motivation to do their own research [\textbf{P143}].

\subsection{Entertainment}
A small, but distinct segment of our sample conceptualized misinformation as any information with a \textit{tangential} faithfulness to known facts relative to humorous, sarcastic, or memetic contexts, usually created and disseminated with intention to ``\textit{entertain}'' [\textbf{P60}] the social media users. Memes form the largest part of the entertaining misinformation, usually containing ``\textbf{erroneous statements}'' [\textbf{P167}] and in some cases ``\textit{making fun of the misinformation itself}'' [\textbf{P174}]. The concept of misinformation as entertainment is not to (counter)argue with posts perceived as polarizing, but to \textit{mock of-the-wall post} [\textbf{P217}] themselves. Inherently fallacious information also contributes to the entertainment, if not mocking, element, for example, ``\textit{celebrity rumors}'' or ``\textit{tabloid content}'' [\textbf{P134}], causing a ``\textit{laughing reaction}'' as well [\textbf{P135}].

\section{Origins of Misinformation}
\subsection{Users from the `Other' Side}
The political counter(argumentation) as misinformation posits a deliberate \textit{selection} of facts that fit into one's political inclinations. Naturally, the conceptualization in this folk model considers  ``\textit{people with different political perspectives wanting their beliefs validated}'' [\textbf{P106}] to be the source of misinformation on social media. These people ``\textit{select facts in accordance with their selfish agendas}'' [\textbf{P117}] and constantly ``\textit{try to convince other people there point of view is correct}'' [\textbf{P122}]. The reason for bringing misinformation to social media is that ``\textit{these people are made in the image of the political leaders they flock to}'' [\textbf{P13}]

The participants in our sample described the users from the `other side' in political counter(argumentation) as: ``\textit{uneducated, bigoted, prevaricating, and shameless hypocrites}'' [\textbf{P66}]; ``\textit{truly deluded, insistent on being stupid}'' [\textbf{P70}]; ``\textit{people who refuse to accept reality}'' [\textbf{P80}]; ``\textit{impulsive, uneducated people}'' [\textbf{97}]; and ``\textit{far-right or far-left activists}'' [\textbf{P101}]. It could be very well that these descriptions refer to the vocal social media users that amplify and spread out-of-context narratives ``cooked in political echo chaimbers'' or ``PACs''  [\textbf{P225}] instead of referring the users themselves as misinformation selectors. Either way, the `other side' was culpable of ``\textit{willing to overlook truth}'' [\textbf{P117}] to ``\textit{off as important and knowledgeable}'' [\textbf{P71}] 

The ``\textit{bitterness} [\textbf{P76}],'' in the mind of our participants ``\textit{comes from people that `know' the information is wrong, but they still like to create drama, so will post it anyway just to see the kind of arguments that will come from it}'' [\textbf{P95}]. The accusations, thus, do not cast the `other side' as ignorant \cite{Kirchner} or lazy \cite{Pennycook1}, but on the contrary as agitated and determined to win an argument. As the participant \textbf{P106} put it: ``\textit{these are people wanting their beliefs validated so they seek out facts that confirm them, then latch onto ones that appear to do so, and sometimes articulate and spread them in a context that turns them into misinformation; I imagine that's more common that outright intentionally telling lies}''. 

\subsection{`Die Hard' Speculators}
In the out-of-context folk model, the interpretation of an issue in an alternative, improbable context is the catalyst that turns a speculative narrative into misinformation. Tracing the genesis back to its source(s), participant \textbf{P189} explained: ``\textit{Misinformation comes, I think often, through selective reading. People want to confirm their narrative, and so they take things out of context, or in limited context. In reality, things are usually more complex. But rather than deal with complexity, simplistic takes that confirm pre-existing narrative biases get read (and shared) more on social media.}'' Several others joined to characterize these people as ``\textit{well-intentioned, but ignorant}'' [\textbf{P210}] or ``\textit{well-meaning dummies}'' [\textbf{P214}]

Misinformation often originated, according to our participants, from ``\textit{twisting what's actually a personal opinion into one's subjective idea of a fact}'' [\textbf{P201}] or ``\textit{expressing one's emotions as facts (speaking a personal `truth')}'' [\textbf{P190}]. This alone might not be sufficient for misinformation to float on social media, but the affordances of the platforms encourage ``\textit{innocent garbling, like the `telephone' game as kids}'' [\textbf{P172}] make the perfect conditions for mutation and further propagation of individual speculative narratives. Participants described this `vicious cycle' starting when ``\textit{someone tells a second person their opinion. Because the second person see's the first person as credible/expert/authority, they're likely to consolidate it in their memory as truth. Just like the game of telephone, facts get lost over time so a narrative gets misconstrued by mutating the message so many times}'' [\textbf{P154, P159, P213, P230}]. 

Many participants accused the `die hard speculators' [\textbf{P85}] as ``\textit{gullible}'' [\textbf{P163}], ``\textit{mentally unbalanced}'' [\textbf{P131}], ``\textit{having poor critical thinking skills}'' [\textbf{P178}], ``\textit{idiots that can't use their brains}'' [\textbf{P185}], or ``\textit{being brainwashed}'' [\textbf{184}]. Some of them are simply ``craving attention'' [\textbf{P182, P133, P202}], some ``just want hear themselves talk''  [\textbf{P179}], some have ``cognitive dissonance'' [\textbf{P202}], and some are ``suspicious of everything due to fear, distrust in science, or living in an `echo chamber''' [\textbf{P168, P188, P221}]. The presumption of innocence is maintained by our participants who think of these social media users as ``\textit{normal people who sincerely believe everything they read, not some big bot, troll campaign, or the work of a foreign actor}'' [\textbf{P146, P160}]. 

Participants identified the `die hard speculators' as responsible for misinformation-as-entertainment, as our participants pointed to ``\textit{people making jokes and other people believing them as real}'' [\textbf{P234}]. The ``\textit{pranksters and internet trolls make up memes and speculation for fun}'' [\textbf{P19, P144}] and the social media platforms enable them to ``spreads like wildfire'' [\textbf{P226}], causing misinformation to come to attention to ``\textit{people who are naturally drawn to posts that is really wild and far-fetched}'' [\textbf{P168}]. Due to the lack of agreed context for interpretation \cite{Albright}, users inclined to ``\textit{spread a negative view towards a figure, issue, or movement}'' [\textbf{P10}] turn misinformation-as-entertainment into entertainment-as-misinformation on social media, and ``\textit{an insane joke whispered in the dark is suddenly whisked around the world}'' [\textbf{P47}].

\subsection{Nation-States}
Expectedly, the `usual suspects' of external propaganda -- ``\textit{nation states hostile to the United States and her interests}'' [\textbf{P183}] -- were identified as misinformation originators by our participants. These nation states were accused as willing to ``\textit{destroy our democracy}'' [\textbf{P128}], ``\textit{cause a decline in society}'' [\textbf{P131}], ``\textit{sow chaos}'' [\textbf{P208}] or ``\textit{destabilize the region}'' [\textbf{P28}]. Many of our participants associated misinformation from nation-states as ``\textit{online propaganda}'' [\textbf{P4, P29, P33, P71, P82, P109}], ``\textit{psychological warfare}'' [\textbf{P21}] or a ``\textit{disinformation campaign}'' [\textbf{P61, P101, P129}]. The reference to ``\textit{bots}'' [\textbf{P16, P27, P43, P101, P121, P158, P235}] was far more prevalent than ``\textit{troll farms}'' [\textbf{P33, P80}] and the bots were accused of conspiring with ``\textit{paid shills}'' [\textbf{P28}] for the deliberate posting of ``\textit{wrong facts}'' [\textbf{P29, P84}].

\subsection{Mass (Social) Media}
Next to the `Russian bots,' another prominent sources of external propaganda the ``mass media industry'' [\textbf{P44}] and the ``social media platforms'' themselves [\textbf{P193}], according to our participants. The mass media outlets, with their ``\textit{biased journalism}'' [\textbf{P215}] and ``talk shoes'' [\textbf{P149}, ``\textit{disseminates `dubious' information according to an agenda through articles, stories, and reports that trickle down to ignorant people}'' [\textbf{132}]. It happens on both sides of the political spectrum too, from ``\textit{right-wing think tanks to bourgeoisie controlled liberal media}'' [\textbf{P216, P27}]. The  `non-partisan' media is also involved as ``\textit{it `spins' both left- or right-leaning narratives for own benefit, but don't like to admit it}'' [\textbf{P125}]. 

The proverbial `repetition makes a statement seem more true' was ascribed to the mass media's \textit{perverse tactics of churning overly sensationalizing headlines}'' [\textbf{P116}] and ``ramming down everyone's throats'' [\textbf{P193}]. The sensationalizing was not exclusive to politics, but also pertains to manipulation relative to ``\textit{foreign affairs, US-centric societal issues, illegal immigration, and public health}''  [\textbf{P219, P225}]. The finger was pointed at ``\textit{Fox and Newsmax}'' [\textbf{P103}] as corporations for this `misinformation offense' but Rupert Murdoch was also accused of ``\textit{having similar goals}'' [\textbf{P60}].

``\textit{Leftist run}'' [\textbf{P96}] social media platforms (e.g Twitter, Facebook) produced misinformation by ``block and ban anything they disagree with while amplifying stories they support which are often lies'' [\textbf{P211}].``Right wing'' [\textbf{P13}] social media platforms (e.g. 4chan, Gab) produces misinformation by applying a ``\textit{hostile context}'' [\textbf{P211}] to every issue ``\textit{they are disenchanted with}'' [\textbf{P58}]. No single social media platform is exempted from misinformation creation as ``\textit{the leftist platforms have to censor information and create propaganda, but the rightist platforms also abuse the free speech environment for winning arguments}'' [\textbf{P108, P13}]. Participants also dropped names, deriding Jack Dorsey for ``\textit{locking out damaging information about Biden's son}'' out of Twitter [\textbf{P64}] and Mark Zuckerberg for ``\textit{literally making billions playing all sides against each other}'' [\textbf{P113}].

\subsection{Domestic Establishment}
Some of the participants suspected that ``\textit{trying to blame Russia is just a deflection from other more important issues}'' [\textbf{P15}], and accused the domestic establishment of ``\textit{not being fully forthright with sharing all the facts available at times}'' [\textbf{P124}]. Instead, misinformation is ``\textit{all the nonsense spread by out institutions to keep people compliant}'' [\textbf{P105}]. The power structures and intelligence agencies we perceived by some participants as having a ``\textit{terrible track record of disseminating misinformation on social media}'' [\textbf{P66}] as they ``\textit{pair their Russian counterparts}'' for both ``\textit{political and personal gain}'' [\textbf{P52}], always supported by ``\textit{dark money}'' [\textbf{P151, P66, P33, P164}]. When it came to `dark money' the names of Koch brothers and Peter Thiel came up as originators of misinformation, aiming to  ``\textit{destroy liberal democracy, rule of law, and replace it with a kleptocratic form of government where the financial elite no longer have pesky regulations or taxes}'' [\textbf{P60}].

\section{Misinformation Purpose}
\subsection{Political Ammunition}
Most of our participants thought that misinformation ``\textit{serves political purposes as it incites people to hate political opponents for bogus reasons}'' [\textbf{P21}]. Misinformation, exploiting people's ``\textit{confirmation bias}'' [\textbf{P107}], ``\textit{makes people more politically close minded}'' [\textbf{P214}], out participants pointed out. By hardens simplistic viewpoints, participant \textbf{P190} reckons that ``\textit{misinformation reduces complex phenomena to sound bites. It has reduced civic discourse in this country to two competing (inaccurate) camps that think the other side is either evil or dumb}.'' Misinformation, participants \textbf{P44} and \textbf{P213} add, ``\textit{creates a `hive mind' on social media from a group of followers that becomes loyal to a political cause and amplifies the sounds bites}''

Our participants also ushered direct accusations
against particular political points of contention. One group of participants believed that misinformation's purpose is ``\textit{to protect democrat politicians}'' [\textbf{P95}] ``\textit{as repetition of lies and misinformation is a very old liberal tactic of brainwashing the populace}'' [\textbf{P156}]. Another believed that misinformation's purpose is to \textit{harass liberals}'' [\textit{P104}] and \textit{convince people to vote for Donald Trump} [\textbf{P31}]. If this was in the past, in the present, some of the participants believe misinformation \textit{keeps the conservatives occupied, encourages discourse}'' [\textbf{P211}]. But, in equal measure, misinformation ``\textit{creates the illusion that left-wing positions are popular}'' [\textbf{P68, P109}]. Overall, an impression holds that ``\textit{misinformation serves to foster division between ideologies, prop up `straw man' arguments, and advocate for particular legal and judicial outcomes}'' [\textbf{P32}].

\subsection{`Stirring the Pot'}
A prevalent conceptualization of misinformation's purpose within our sample was to ``\textit{stir the pot}'' [\textbf{P7}] and ``\textit{bring some sort of anarchy or civil disobedience}'' [\textbf{P151}]. Along these lines, misinformation's purpose was described as: ``\textit{to suppress class consciousness} [\textbf{P28}], for ``\textit{muddying the waters}'' [\textbf{P4}], to ``\textit{keep people up in arms}'' [\textbf{P13}], to wreak ``\textit{havoc in the society}'' [\textbf{P16}], ``\textit{destabilize the nation}'' [\textbf{P26}], and ``\textit{absolutely destroy America and democracies around the world}'' [\textbf{P60, P81}]. Fear was pointed out as one of the `ingredients' for ``\textit{leading people on and thus creating confusion about the truth}'' [\textbf{P20, P82}]. Scaring with misinformation was the best tactic for `` political obedience'' [\textbf{P38}], ``riling up people to fight shadows'' [\textbf{P51}], ``\textit{diminish trust in each other}'' [\textbf{P75}], and ``\textit{eating up the culture war pretext}'' [\textbf{P29}]. 

Confusion and doubt we naturally included too, as misinformation was seen to be the ``\textit{ammunition for being contrary, either to subvert existing structures or get some sort of satisfaction out of being obnoxious}'' [\textbf{P79, P130}]. The goal of misinformation, thus, was driven to ``\textit{create less trust in institutions}'' [\textbf{P183, P170, P69}] ``\textit{distraction from real issues}'' [\textbf{P124, P11, P46}], and  ``\textit{directing the consensus to a topic based on hidden agendas}'' [\textbf{P142, P86, P14}]. Several participants took misinformation to be the main tool for ``indoctrination'' [\textbf{P38}] that ultimately results in ``killing Americans'' [\textbf{P207, P20}]. Few thought of misinformation's purpose as a ``\textit{laboratory experiment}'' on a massive scale conducted both by the US military (public opinion manipulation) [\textbf{P112}] or foreign directorates of `active measures' (weakening US economy and military power without fighting a real battle.) [\textbf{P206}].

\subsection{Profit}
Monetary profit was an equally relevant conceptualization of the misinformation's purpose within our sample. As participant \textbf{P176} describe it: ``\textit{Misinformation drives page views; Page views generate money; Follow the money}.'' Facebook was explicitly singled out as ``\textit{promoting misinformation posts to boost engagement and ad revenue on their platform}'' [\textbf{P70, P26, P140, P83}]. Misinformation ``\textit{increases use and readership of the social media}'' [\textbf{P166, P150, P63}] and thus it is attractive for ``\textit{clickbait campaigns, selling products, merchandise, or garner donations}'' [\textbf{P39, P50, P148}] as well as to ``pump the price of certain cryptocurrencies'' [\textbf{P27}]. Being able to ``\textit{confuse and/or upset people}'' [\textbf{P83}], ``\textit{misinformation ultimately is a for-profit enterprise}'' [\textbf{P39}], for ``\textit{mass media profiteering}'' [\textbf{P104}] or ``\textit{driving stocks prices}'' [\textbf{P20}].

\subsection{Entertainment}
A subset of our participants believe that misinformation ``\textit{doesn't have a true function besides satire and entertainment}'' [\textbf{P80}]. Therefore, a perception holds that ``\textit{it is entertaining to see who believes the crazy stuff that is out there}'' [\textbf{p14}]. In the words of participant \textbf{P130} misinformation is a source of entertainment because ``\textit{it is fun to watch people describe daily all sorts of impossible stuff that is bothering them}.'' The attention seeking element was also mentioned by our participants, suspecting that misinformation serves a purpose to make ``\textit{\textit{users on social media look cool or funny}}'' [\textbf{P192}].


\section{Assessing Misinformation}
\subsection{Analytical Assessment}
We found a strong presence of the dual model of online information's credibility assessment suggested in \cite{Metzger, Pennycook}. Many of our participants combined both analytical skills and heuristics in determining whether a post is misinformation on social media. Participant \textbf{P8} offered a simple rule of thumb: ``\textit{If it seems insane that is the first clue. Beyond that, I look at the site or the source being used. If there is no source? 90\% chance it's a lie. If there is a source/site listed it doesn't take much effort to glance at it and know if it's misinformation or an extremist site}.'' Relying on the political counter(argumentation) folk model, our participants reasoned that ``\textit{in a political discussion, especially an argumentative one involving more than one perspective, it's likely that at least some misinformation is being spread, and this is often self-evident when one side makes one claim and another makes the opposite, as both claim and counter claim can't be true at the same time}'' [\textbf{P15, P110, P140, P156, P192}].

When it came to the out-of-context narrative folk model, participants were well versed in  differentiating between contexts and knowing how often, within a context, misinformation is likely to be spread. Any posts that ``\textit{contain cherry picking events and presenting things out of context in order to support their biased arguments} were clearly a misinformation for Participants \textbf{P28}, \textbf{P34}, \textbf{P90}, and \textbf{115}. For these participants, it was easy to ``\textit{look for the main context from other reputable sources}'' and reject such posts as misinformation. Usually, if a post's ``\textit{sole purpose seems to be to evoke at knee-jerk response}'' [\textbf{P60}], participants went for a third party confirmation ``\textit{from a source with journalistic integrity}'' [\textbf{P58}]. Here, the rule of thumb, as participants \textbf{P39}, \textbf{P55}, \textbf{P21}, and \textbf{P119} pointed out, was: \textit{``If it's a share or post of an article, I look at the source.  Is it a news organization, and if so, which one? Or is it a blog? Is the article using `trigger words,' hostile language, or blanket statements? If yes, most probably is misinformation''}.

Interestingly, there was a division on the question of journalistic integrity (or lack of thereof) as one group of participants strictly ``\textit{stayed away from anything from Fox news, Beritbart, or outlets I never heard of}'' [\textbf{P173, P99, P24, P59, P84}] and another group from the ``\textit{liberal scare tactics of New York Times, NBC, CNN, and Washington Post}'' [\textbf{P144, P156, P152, P216, P230}]. This, together with the notions of the ``\textit{external boogeyman accounts of the Russians containing spelling mistakes, bad grammar, and weird patriotic sounding name}'' [\textbf{P225, P230, P173, P171}] conducted their assessment employing the external propaganda folk model of misinformation. Participants employing the inherently fallacious information folk model resorted to using ``\textit{fact checking sites like PolitiFact.org and Snops}'' [\textbf{P9, P76}], after they initially went ``with their gut'' [\textbf{P187}]. Several of our participant then turned back to social media with the fact-checked version of the issue at stake, and decided that definitely is a problem of misinformation when the ``\textit{poster stopped responding when challenged to provide sources or refute the fact-checked version, they stop responding}'' [\textbf{P55, P66, P118}].  

A cue that threw several participants off about misinformation is when posters ``\textit{started throwing random polls or numbers out there and don't talk about where they come from}'' [\textbf{P137, P30, P44}]. For these participants, the rule of thumb was to turn to ``\textit{scientific evidence}'' [\textbf{P174, P219, P47}] beyond the fact-checking websites as the felt ``\textit{confident to find the truth after years of fine-tuned bullshit detection}'' on social media [\textbf{P135}]. Their go-to approach was to ``\textit{presume false, until proven true}'' [\textbf{P213}] when posts ``\textit{seem to be written to invoke an immediate response}'' [\textbf{P142}]. Few participants even critiqued the reliance only on secondary fact-checking, as ``\textit{people count on mainstream media to tell them what is true or false, and that is wrong}'' [\textbf{P214}]. Everyone on social media, in their view ``\textit{should take a piece of information and do background on it themselves to find out what is true and what is fake}'' [\textbf{P149, P157, P104}].

A counter evidence, albeit anecdotal, to the line of research suggesting that people, often times do not heed social media misinformation warnings \cite{Sharevski-cose, Clayton, Nyhan, Ecker, Thorson} (with one exception, participant \textbf{P66} noted that ``\textit{sometimes misinformation is labeled as such; that is not always accurate though.}''). Participants \textbf{P19}, \textbf{P95}, \textbf{P167} \textbf{P208}, for example, noted if a post has ``\textit{a thing that the whatever has been fact checked by Facebook of Twitter and is questionable}'' then they did their own research to confirm it as a misinformation. Participants \textbf{P128}, and \textbf{P55} went further and said: ``\textit{sometime posts have been flagged as misinformation by the platform, which is very helpful; I come to realize this happens usually for topics that seems divisive or extreme, so these two elements are my key indicators to misinformation on social media}''. Participant \textbf{P221} and \textbf{P231} added that they look up the ``\textit{flagged posts for misinformation by the social media}'' on Google to ``\textit{dig deep}'' and do research to get to the bottom of the misinformation issue.

\subsection{Heuristics Assessment}
Prior evidence on heuristics-based assessment of misinformation online suggests that people either look for a cue at the combination of news title plus the image appeal \cite{Spezzano} or the the credibility of the account/liked website \cite{Kirchner}. We found a confirming evidence for both cues in our sample, but with a much more nuanced understanding of how social media users' apply this in their daily lives. On the first heuristic cue, participants stated that ``\textit{you can tell by whether the headline sounds implausible or misleading}'' [\textbf{P27, P66}], posts are ``\textit{just meme-type images saying something controversial}'' [\textbf{P85, P9, P59}], or posts are coming from news outlets that constantly ``\textit{demonize the other side}'' [\textbf{P58}]. Employing both the political counter(argumentation) and the out-of-context narrative folk models, participants repeatedly used to describe this cue as ``\textit{outlandish}'' in reference to both the textual and audiovisual content of such posts that usually revolve around ``politics, health, celebrities, or conspiracies [\textbf{P89, P109, P115, P199. P224}]. 

Here, we noticed that when participants referred to outlandish posts as misinformation, they pointed usually to social media posts about which they were largely unfamiliar with, or had only partial evidence from doing a research on their own. When they felt confident and very familiar on the issue, we uncovered a cue we named ``\textit{too good to be true}'' as multiple participants described their go-to misinformation detector on social media as such [\textbf{P180, P197, P200, P209, P223, P18, P35, P43, P75, P90, P96, P142, P150}]. Referring to this cue as the ``\textit{misinformation red flag}'' [\textbf{P102, }] participants thought that ``if it looks too good to be true, it probably is'' when encountering posts that ``\textit{conflict with real life}'' [\textbf{P112, P77, P92, P11}] or simply are against the ``\textit{common sense}'' [\textbf{P106, P122, P123, P132, P176, P177, P186, P230, P9}]. 

When it comes to the ``\textit{sourcing}'' cues, participants used an extended set of ``\textit{gut feelings}'' [\textbf{P213}] to figure out who is the `misinformer' on social media. If the user name has a ``\textit{bunch of numbers, than almost certainly is a bot account spreading misinformation}'' [\textbf{P100, P162, P124}]. Participants \textbf{P50} explained that these accounts ``\textit{have some generic term in the user name next to the numbers, have very few followers and their bio has a link to a website that is clearly not legitimate}.'' If language in the post had ``misspellings'' that was the first strike; if the ``grammar was attrocious'' that was the second strike; and if the post had a ``far-fetched context'' that was the third-strike-you-are-misinformation rule employed by participants \textbf{P123} and \textbf{P171}. The ``\textit{poorly written}'' cue  was sufficient for a lot of the participants as a ``telltale'' sign of misinformation as the grammatical, syntactical, and linguistic mistakes were often accompanied  with ``\textit{outdated aesthetics}'' or ``\textit{memes contain a message that otherwise will be censored}''  [\textbf{P69, P123, P32, P107}]. 

The ``\textit{emotion-check}'' cue, we found out, was frequently employed by the  participants in our sample to discriminate between misinformation and content faithful to known facts. As participant \textbf{P86} put it, ``\textit{if something gets an emotional reaction out of you, it is time to question the veracity}.'' Emotion-provoking misinformation, per our participants \textbf{P4}, \textbf{P94}, and \textbf{P36}, emerges on social media when posts ``\textit{speak in absolutes and employ either absurd or illogical claims}.'' According to participant \textbf{P29}, if ``\textit{the source alone is not a sufficient cue, than, misinformation is easy to pick out because it uses anger, fear, or malice to get the point across}''. Usually, the emotion-provoking misinformation \textit{consistently speaks about one thing or have very strong opinionated comments beneath about the same thing}'' [\textbf{P131}].

\section{Responding to Misinformation}
Users on social media employ various tactics when responding to social media. The main actions taken as part of a tactic involve: i) fact/source checking, ii) blocking, muting, unfollowing, unfriending, iii) reporting the post/account, iv) countering with replying/commenting, or v) noticing but ignoring. From our sample, we identified the following tactics considering a user encountered and read a misinformation post on social media: 

\begin{enumerate}
    \item  fact check $\rightarrow$ ignore
    \item  fact check $\rightarrow$ block/mute account 
    \item  fact check  $\rightarrow$ report
    \item  source check $\rightarrow$ fact check $\rightarrow$ report
    \item  fact check $\rightarrow$ public refute
    \item  fact check $\rightarrow$ repost accurate sources $\rightarrow$ overshadow misinformation with information  
    \item block/mute (3-strikes-out) $\rightarrow$ ask others to block/mute
    \item block/mute $\rightarrow$ report
    \item report $\rightarrow$ only when deemed `dangerous' misinformation
    \item report $\rightarrow$ disengage from platform
    \item reply $\rightarrow$ report
    \item reply $\rightarrow$ detailed disproving with facts
    \item reply $\rightarrow$ challenge the account to debate
    \item reply $\rightarrow$ use the post to educate others on the misinformation
    \item ignore $\rightarrow$ fear of retribution 
    \item ignore $\rightarrow$ give it a day for the story to develop $\rightarrow$ respond to refute
    \item ignore $\rightarrow$ treat them like spam or junk mail
    \item ignore $\rightarrow$ expect the platform to remove it
    \item ignore $\rightarrow$ tell my inner circle (comment within)
    \item ignore $\rightarrow$ privately refute only if it's someone close to me
        \item ignore $\rightarrow$ laugh about it
\end{enumerate}

\subsection{Fact/Source Checking}
Fact checking a post was a tactic that most participants employed when they ``\textit{actually cared about the issue}'' [\textbf{P63}]. As an analytical assessment, participants took various ways to actually do the fact checking. One group resorted to fact checking websites like ``\textit{Snopes or FackChecker.org}'' [\textbf{P75, P134, P167}], another went ``\textit{`googling' the issue}'' [\textbf{P155, P175, P180, P221}], and a third used only ``\textit{reputable sources or official reports}'' [\textbf{P40, P199, P167}]. A small subset opted for a balanced fact checking, reading ``\textit{both left- and right-leaning sources}'' in addition, and looking for the ``\textit{truth somewhere in the middle as journalists are inherently biased}'' [\textbf{P150, P191}]. 

The decision to take any action or just ignore the post comes after a brief period of ``\textit{critical thinking}'' [\textbf{P63}] when the participants ``\textit{pick the battles}'' [\textbf{P143}] considered the account behind the post and the ``\textit{`outlandishness' of the narrative}'' [\textbf{P129, P146}]. The `rule of thumb' within our sample was to \textbf{\textit{fact check, but ignore}} if the the account is well known ``\textit{misinformation spreader}'' [\textbf{P130}] and the post was ``\textit{unapologetically fabricated}'' [\textbf{P120}]. When taking an action, our participants resorted to \textbf{\textit{fact check, then block/mute the account}} when they deemed ``not worthwhile to try and argue with them because they will not listen'' [\textbf{P38, P44}], but still prudent to do their ``own research'' [\textbf{P154, P7}]. The \textbf{\textit{fact check, then report the account}} tactic was employed if the participants were ``convinced that the account(s) are hostile actors from other countries, bots, or intentionally grifting'' [\textbf{P51}]. Some of them did a preemptive \textit{\textbf{source check}}  ``\textit{looking for grammatical errors, the website URL and citations for known facts, before reporting the post}'' [\textbf{p217}]

A group of participants showed a proactive stance when responding to misinformation on social media as they employed a \textit{\textbf{fact check, then engage in public refute}} tactic [\textbf{P13, P136, P148}]. They looked for ``\textit{ facts, numbers, charts, quotes---anything that will refute the misinformation items stated}'' [\textbf{P201}]. Some of these participants took an interesting approach of \textit{\textbf{overshadowing misinformation with real information}} as they resorted to ``\textit{flood the poster with re-posts, tags, and and replies}'' including accurate sources [\textbf{P222, P165}]. In their view, showing a direct resistance to the misinformation spreaders by ``publicly calling them out'' [\textbf{P101}] is equally important and necessary as it is the ``\textit{misinformation cleanup}'' [\textbf{P29}] the social media platforms should do a better job of. 

\subsection{Blocking, Muting, Reporting}
Blocking, muting, and reporting without fact checking, but instead based on subjective convictions or heuristic assessment, was also a regular tactic employed in our sample. Some of the participants were quite lenient, giving a \textbf{\textit{three-strikes-you-are-out}} chance for someone on social media to post what they perceived as misinformation before they proceed to ``\textit{block, mute, or un-follow and encourage others to do so}'' [\textbf{P72, P192, P39, P51, P81}]. \textbf{\textit{Block/mute, then report}} was a step further taken by participants in our sample, in hope ``\textit{the account and the post are taken down}'' [\textbf{P22, P30, P10}]. These participants avoided commenting ``\textit{lest I accidentally help amplify the post}'' [\textbf{P27, P58, P30, P55}]. 

Some of the participants took an approach where \textit{\textbf{misinformation was only reported when deemed `dangerous'}}. The reporting was directly to social media admins using the platform affordances. Under `dangerous' our participants considered any misinformation that is ``\textit{threat to public health},'' ``\textit{civil unrest},'' or \textit{defamatory in nature}'' [\textbf{P196, P33, P39, P50, P55, P114}]. Some of the participants noted that they, ``\textit{\textit{after a repeated exposure}},'' have decided to simply disengage with a social media platform after giving its admins full hands of work by ``\textit{reporting a list of misinformation spreaders}'' [\textbf{p112, p53}]. In their eyes, social media became a ``\textit{garden too hard to weed from misinformation}'' [\textbf{P83, P79, P91, P94, P144}].

\subsection{Replying}
Our participants in a large degree put both their analytical heuristic assessment skills in directly engaging with the misinformation spreaders on social media. Encountering ``\textit{inflammatory rhetoric}'' [\textbf{P142}], they first \textbf{\textit{replied, then reported}} it, ``\textit{leaving a comment explaining why this rhetoric it is misinformation} [\textbf{P200, P165, P144}]. Many of them skipped the reporting but invested in \textit{\textbf{detailed disproving with facts}} by ``\textit{advising the person who posted it that the information is blatantly incorrect and what the correct information is with supporting documentation}'' [\textbf{P202, P150}]. Participant \textbf{P182} reasoned that misinformation spreaders ``\textit{'deserve' ad hominem counter-argumentation fusillade dripping with vitriol}'' in case ``\textit{disproving the the misinformation with facts is futile.}''

Several participants called on the `misinformer' to ``\textit{remove the post}'' [\textbf{P24, P15, P142}] afraid of its inciting a massive rift. Directly \textbf{\textit{challenging the account}} spreading misinformation by ``\textit{engaging in a reasoned debate}'' [\textbf{P46}] was also a tactic where participants judged that a constructive discourse is possible. Usually, the reply includes ``\textit{truthful statements that challenge their misinformation}'' [\textbf{P22, P4, P14}]. Sometimes, the counter-argumentation was framed in a ``shoe on the other foot'' metaphor. For example, participant \textbf{P90} shared that they ``\textit{reply with something to make the misinformer think about what they're doing (e.g. what if this is your mom) and the repercussions they might bring to someone.}''  

The presences of `misinformation spreaders' in participants' circles on social media enabled the participants to engage with the for the purpose of \textit{\textbf{educating the others about the perils of misinformation}}. Participant \textbf{P29} indicated that they ``\textit{will reply to the person, not trying to convince them of course but rather to help educate others who might be reading the misinformation}'' as to strategy to ``\textit{neutralize the spread of misinformation}'' [\textbf{P176}]. In the view of participant \textbf{P32} this tactic ``\textit{counters misinformation with as much sourced, unbiased information as possible to the benefit so others could see how easy is to disprove misinformation}.'' Our participants are aware that arguing with the `misinformer' leads to ``\textit{frustration}'' [\textbf{P145, P232, P193}] but is worthwhile because it helps participants themselves ``\textit{to talk to real people about how to counter it}'' [\textbf{P107}].

\subsection{Ignoring}
Many participants in our sample simply indicate they ``\textit{just ignore misinformation}'' [\textbf{P187, P215, P229, P2}]. One reason, as participant \textbf{P188} indicated, is because ``\textit{social media became largely an ineffective place for political discourse, so there is no point to engage with misinformation}.'' Another reason is that participants were \textit{\textbf{fearing retribution}} either from ``\textit{the platform as they could ban me for refuting it}'' [\textbf{P220, P3}] or ``\textit{being cyber-stalked}'' by the `misinformer' [\textbf{P18}]. Sometimes participants ignored misinformation but only temporary as they waited for the \textit{\textbf{misinformation story to develop}}. Participants \textbf{P216} and \textbf{P162} pointed out that this is a helpful tactic as it helps to ``\textit{see how much others will refute the post and that itself is a sufficient enough factor to differentiate between misinformation and the truth}.``

An interesting approach for ignoring misinformation on social media was that it simply should be \textit{\textbf{treated as a spam or junk mail}}. There is so much misinformation that one `` \textit{becomes to perceive it like mail that ends in spam or obnoxious digital ads and eventually just stop noticing it}'' [\textbf{P35, P104, P142}]. As such, participants expect that it is the job of the \textit{\textbf{platform to remove misinformation}}, not ``\textit{their problem}'' [\textbf{P157, P26, P29}]. Those still bothered with it, resorted to two extra steps: either \textit{\textbf{took it within their inner circles to comment on}} or \textit{\textbf{reached out privately to the `misinformer' only if they were close to them}}. The first tactic was to shield participants themselves from ``\textit{hateful responses}'' [\textbf{P158, P27}] or avoid ``\textit{amplifying the misinformation further} [\textbf{P75}]. The second tactic, as participants \textbf{P142} and \textbf{P177} noted, is useful as it plays on the personal trust to ``\textit{dissuade the `misinformer' from propagating the chaos}.'' Finally, there were ones that did ignore it but still squeezed a \textit{\textbf{laugh about it}}, in accordance to the misinformation-as-entertainment folk model. ``\textit{Fake},'' for these participants means ``\textit{laugh and move on}'' [\textbf{P156, P9, P25, P168}]. 


\section{Resilience to Misinformation}
Individual resilience is the ability to continue on, even in the face of adversity, and maintain the well-being of oneself \cite{Jurgens}. In the context of online interactions, resilience is understood as the capacity of groups of people bound together in an online community to sustain and advance their well-being in the face of challenges to it \cite{Hall}. The second definition conceives resilience as a collective characteristic that transcends the individual level, and as such, individual adversities are abstracted to structural challenges. Based on this abstraction, \cite{Benkler} studied the resilience to misinformation for media systems in the US, arguing that resilient media systems: a) prevent emergence of a large audience that no longer expects true reporting from its preferred ideological media, but primarily identity-confirming news and opinions—regardless of the truth content; and b) brokers news by applying the principle of accountable verification to all information within the media environment.

Utilizing this analytical approach, \cite{Humprecht} generalized the resilience to misinformation as a structural context in which disinformation does not reach a large number of citizens. And even if it does, \cite{Humprecht} argues, people should be less inclined to support or further distribute such low-quality information, and in some cases, they should be more able to counter that information. This structural context encompasses factors such as: populist communication, societal polarization, trust in news media, social media news consumption, exposure to online disinformation. Testing several countries' information environment on a general level, \cite{Humprecht} concludes that United States is the \textit{least resilient} country to misinformation among the western democracies.  

While both analyses in \cite{Humprecht, Benkler} are useful for general policy-making decision, they fail to consider the individual resilience of the people that are \textit{actually} exposed and engage with misinformation. Seeing misinformation on a group level through the prism of structural factors operates on the assumptions that individuals in online communities and social networks are incompetent \cite{Torres}, lazy (or unwilling)  \cite{Pennycook1}, and deceptively biased \cite{Luo} to withstand the challenge of misinformation. As such, it ignores what folk models are employed to conceptualize misinformation in the first place, as well as how these users respond, internalize, and live with misinformation adversities to preserve their well-beings. This omission, in our view, perpetuates a rather distorted perception that resilience to misinformation is exclusively a problem of external ``inoculation'' against information-based threats \cite{Lewandowsky}. In this top-down approach, media systems should report only news with the highest faithfulness to known facts \cite{Hardy}, social media platforms should label, remove, and ban any inaccurate content \cite{Pennycook}, and social media users taught how to deal with real-world misinformation through exposure to weakened doses of misinformation \cite{Maertens}. 

Our study reveals that social media users in the US have already developed a ``natural immunity'' and are ``self-inoculated'' against misinformation on social media, refuting the top-down approach for misinformation attrition prescribed through group abstractions. Our participants used their mental models to determine whether information is qualified as ``misinformation'' (a threat) or not, take in, process it and ``bounce back'' with participation in the social media discourse (i.e. an individual exposed to misinformation had no adverse reaction to the encounter). The folk models in our study may not be entirely accurate or in full accordance with the prescriptive ``pro-truth pledge'' \cite{Tsipursky} but they return the individual to a well-being state, one in which their version of dealing with misinformation remains true. To show that bottom-up approach to building resilience is equally relevant strategy for the ``war on misinformation'', we contextualize the results in both individual and group ``immunity'' settings.

\subsection{Folk Models of the Self}
\subsubsection{Political (Counter)argumentation Folk Model}
This folk model primarily employs the ignoring tactics as their greatest skill in overcoming misinformation. Seeing political arguments as misinformation, participants opted to ignore the ``other side'' when sensing that either ``\textit{mocking and taunting is about to replace a civilized conversation}'' [\textbf{P220}] or \textit{``the platform will impose it's own abridged definition of `free speech'''} [\textbf{P45}]. However, participants with this folk model switched tactics to reporting if the political arguments was seen as ``\textit{harmful}'' \textbf{[P172]} or ``\textit{if there's a chance of actual harm coming resulting from people believing the misinformation}'' \textbf{[P176]}. The fact-checking tactics were also employed by several of the participants, but mostly driven towards resolving any possible doubts. As participants \textbf{P109} and \textbf{P6} pointed out, they go about `immunization'  ``\textit{depending on the source and type of claim(s)---either investigate the truth of the claim or just dismiss it as the creation of a sick mind}.'' 
 
Blocking/muting and asking others to do so was a tactic justified by self-preservation means of ``\textit{avoiding misinformation trash to not take up space on my timeline}'' [\textbf{P11, P171}], ``\textit{not wasting my time reading it}'' [\textbf{P29, P79}], or ``\textit{not amplifying the other agenda by engaging with it}'' [\textbf{P202, P164}]. Reporting, in the true sense of bottom-up resilience, was seen by the participants employing this folk model as ``\textit{helping the administrators learn the latest lies on the political spectrum}'' [\textbf{P64, P86}], ``\textit{prevent calls for violence, defamation, and hate to materialize}'' [\textbf{P39, P50}], and ``\textit{stop any amplifying bots}'' [\textbf{P96}]. When queried on other ways these participants process misinformation despite blocking, muting, reporting and ignoring it on social media, most of our participants with this folk model said they do in fact talk about it offline within their inner circles. 

Participants \textbf{P57}, \textbf{P51}, and \textbf{P164} underlined that ``\textit{it is important to be aware of the false information that is spreading to be informed of what some others may be thinking, as being ignorant could lead to another January 6th}''.  Many of the ones employing the political (counter)argumentation folk model found it critically important to talk to their parents, grandparents, in-laws, or senior family members ``\textit{point them to the facts and tell them they should stop following/ listening to those people on social media}'' [\textbf{P152, P110, P80, P181}]. Naivety was the vulnerability identified for those social media users lacking extensive heuristics for misinformation assessment or lack of motivation for ``\textit{proactive analysis of source credibility}'' [\textbf{P62}]. Expectedly, several of them switched to the entertainment folk model of misinformation when discussing it offline, to ``\textit{salvage a good laugh from the twisted minds of the political zealots}'' [\textbf{P182, P15, P159, P1}]. 

\subsubsection{Out-of-Context Narrative Folk Model}
The participants operating with this folk model were equally keen on fact checking and engaging with the ``misinformers'' on social media as their ``\textit{pledge to refute false assertions}'' \textbf{[P148]}. Bringing facts in a widely accepted context was more important than simply blocking ``misinformers'' because, as participant \textbf{P174} reasoned, ``\textit{engaging with misinformation posts helps understand where the alternative context comes from and who is behind it}.'' Participant \textbf{P178} strengthened this position asserting that ``\textit{it is important to help educate those that listen to such posts; turning away at least one individual is worth the cost in confronting any deluded `misinformers'}'' Out-of-context narratives, if not directly confronted, ``\textit{turn social media into a breeding ground for hatred, racism, misogyny, greed, and wild conspiracies}''  participants \textbf{P197}, \textbf{P158}, \textbf{P127}, and \textbf{P30} believe.  

Participants with this folk model also possess a considerable degree of self-preservation of their well-being. For example, participant \textbf{P36} explained: \textit{``I have blocked more people since COVID-19 started than the 10 years I have been using Facebook. Before COVID-19 I had blocked 3 people. From the start of COVID-19, up until now, I have blocked over 20 people speculating nonsense to a point of harassment''}. Participant \textbf{P8} added: ``\textit{block the people who post misinformation move on - this is the only way to combat it and remain somewhat sane}.'' The ignoring set of tactics helped users with the out-of-context folk model to bounce back from the lack of control they felt. Participant \textbf{P68} noted: ``\textit{I read them, and move on. I can't do anything to stop them}.'' Continuing on unbothered was premised on the resilience idea that ``\textit{the `misinformers' come to social media to be acknowledged as such}'' [\textbf{P12}]; most of the misinformers welcome push back not just because they are ready to converse, but because that ``\textit{confirms their `misinformer' identity}'' [\textbf{P69, P24}].  

The out-of-context narrative folk model, unlike the political (counter)argumentation, set some of the participants to reject introducing misinformation in their real life as it was deemed a ``\textit{mood killer}'' or a ``\textit{unnecessary cause to our already high stress level}'' [\textbf{P8, P9, P176, P178}]. Another reason was to avoid ``\textit{making enemies among friends and acquaintances based on difference in interpretation of hot topics}'' [\textbf{P176, P187, P53, P33}]. The urge for education, especially of senior family members was also present here, especially because they were the categories that ``\textit{early in the COVID-10 pandemic were ignoring the mandates and potentially spreading or dying from the virus}'' [\textbf{P142, P176}]. Here too, the misinformation-as-entertainment model was assumed in private offline interactions where participants ``\textit{held their comments until amongst the friends to laugh together}'' [\textbf{P130}].

\subsubsection{Inherently Fallacious Information Folk Model}
The participants operating with the inherently fallacious information model look at misinformation as something that can be thwarted by direct action on social media platforms. Blocking and reporting therefore were utilized to flag users and misinformation posts to the attention of the platform administrators. As participants \textbf{P54} and \textbf{P20} put it: ``\textit{We do not share them, we report them to site administration and leave a comment stating they are not legitimate and hope they are taken down}.'' Participants \textbf{P28}. \textbf{P57}, \textbf{P142}, and \textbf{P34} added that they ``\textit{use social media to consume useful information, and don't want their feeds or the feeds of others to be polluted with garbage content}''  Participants in this folk model refrained from complaining about the futility of reporting or in failures of social media platforms in managing misinformation. 

In the view of participants \textbf{P163} and \textbf{P196}, the blocking and reporting should be done proactively to ``\textit{counter accounts that are often posting misinformation}'' so the spread is contained as soon as possible. While some of the participants noted they, by now, were ``\textit{desensitized to misinformation}'' [\textbf{P234, P99}], the sense of hopelessness and discouragement to directly engage with misinformation was considerably less prevalent than among the other folk models. On the contrary, the engagement was seen with a purpose as these participants employed the three-strikes-you-are blocked/muted strategy the most. Equally, participants employed the reply and report in an effort ``\textit{to help educate others who might be reading the misinformation}'' \textbf{[P47]}.

\subsubsection{External Propaganda Folk Model}
Seeing misinformation as an external interference, or simply propaganda, set a subgroup of participants to action as means to counter persuasion. As propaganda messages could  fluctuate in regards the faithfulness to known facts, these participants prioritized proactive fact checking to ``\textit{overshadow it with information with maximum faithfulness to known facts}'' [\textbf{P129}]. Part of this tactic was demonstrating resistance to ``misinformers'' and part was educating people in the inner circles, as participant \textbf{P80} put it: ``\textit{If I have someone around me who is believing the misinformation about specific hot topics at the moment, I will give them tons of references in hopes I can sway them away from someones' hidden agenda}.'' 

While the other folk models employed the dual processing of misinformation in equal degree, the participants using this folk model heavily relied on heuristic processing as their ``\textit{guts}'' \textbf{[P33]} or``\textit{common sense}'' \textbf{[P31]} was telling them ``\textit{not to take emotion-provoking posts at the face value}'' [\textbf{P80}]. The rejections of emotion-provoking content as a resilience strategy worked for these participants are they were confident in debunking ``Russian propaganda'' in particular,  ``\textit{from Pizzagate to denazification of Ukraine}'' [\textbf{P120, P182}]. Some of the participants realized that misinformation, in the context of persuasion, could also hinge on religious and political beliefs and felt that experience-based argumentation was needed, in addition to facts, to counter ``domestic propaganda'' in the form of ``\textit{`don't say gay' laws, anti-abortion laws, or gun laws}'' advocacy [\textbf{P94, P80, P135}]. 

\subsubsection{Entertainment Folk Model}
The misinformation as a form of entertainment folk model primarily ignores misinformation because participants feel they are powerless against it as ``\textit{there is nothing they can do to change the minds of the spreaders and believers}'' [\textbf{P232}]. Participants had a pessimistic view of their ability to affect change in the community because ``\textit{news media can be so confusing, it is hard to know the truth anyway - who I am to tell them that the news is fake}'' [\textbf{P11}]]. They felt that ``\textit{there's no use in reporting it since others probably have already done so gullible people will just continue to spread it until it is removed from the platform}'' [\textbf{P7}]. 

While the laughing at misinformation came from a place of self-preservation (``\textit{it's not worth arguing with them, only laugh at them}'' [\textbf{P12}]), these participants also appear to have an emotional threshold that when crossed, causes them to actively resist by blocking, muting, reporting, or replying. As participants \textbf{P125} and \textbf{84}, ``\textit{these actions are warranted when memes stop being taken humorously and become seeds for more misinformation}.'' These participants were inclined to fact check the memetic accuracy behind misinformation but for their own purposes and shared back to the community (both online and offline) when that threshold is crossed. These appeared to be emotionally driven responses such as ``\textit{I will sometimes mock how stupid the misinformation is and laugh about it openly}'' [\textbf{P125}].

\subsection{Folk Models of a Community}
The ``group'' unit of analysis in \cite{Humprecht, Benkler} largely relies on dichotomous conceptualization of users in pro/against political ideology, mainstream/alternative narratives, or foreign/domestic propaganda camps in discussing resilience to misinformation. While generalizations could be inferred from the responses in our study, the dichotomy seems too reductive in analyzing a bottom-up ``natural immunity'' build-up that synthetizes  conceptualizations of multiple folk models of misinformation. Folk models extend beyond the individual and into communities, especially during polarizing events or threats where self-organization relies on information gathering and cooperation \cite{Jurgens, Kaufmann, Wash-folk}. These threats, for the purpose of our study, were information-based, i.e. threats to the factual, objective, accurate, and honest representation of reality through information posted on social media.   

Past research indicates that an individual in a social media environment will have their  ``resilient (re-)actions mediated, which may not only affect [themselves] and [their] experience, but they also affect a potentially unknown collectivity of users'' \cite{Kaufmann}. In other words, communities on social media could adopt a different folk model, selection of response strategies, and origin/purpose conceptualizations that either heightens or lessens resilience of misinformation. As we have seen from our results, individuals who use the entertainment folk model felt somewhat powerless against misinformation \textit{except when it crosses and emotional threshold}. Those using the political counter(argumentation) folk model had a lower bar to switch from passive to active engagement with misinformation, when misinformation was deemed to have \textit{harmful consequences} both to the discourse on social media and real life (otherwise, the exposure was ``asymptomatic'' as it was done for the purpose of entertainment). And those operating on the inherently fallacious information folk model consciously invested themselves to dissuade users with propagandistic out-of-context narratives folk model on social media. 

Applying this viewpoint to the wickedness of the misinformation problem (usually treated as a vulnerability to misinformation) it becomes apparent that social media communities already started to build resiliency to misinformation without external ``inoculation.'' The definition of resilience in \cite{Benkler} is met in that the individual folk models (and the purpose-based adoption of them on a community level): a) confront, block, mute, and report large audiences inclined to consume identity-confirming news; b) applies the principle of accountable verification to all information within the social media environment through fact checking, research, or consulting scientific evidence. Similarly, the structural context in which disinformation does not reach a large number of citizens \cite{Humprecht} becomes increasingly immaterial as the developing natural immunity to misinformation is able to distinguish between ``selective, questionable, fluctuating, tangential, or absent'' faithfulness to known facts. This is not to say that the folk models alone are sufficient in overcoming the wickedness of misinformation, as the resilience only reduces the effects of misinformation and it does not eliminates or eradicates it. Instead, it is to say that the folk models deserve attention to learn how (mis)information is already processed for the purpose of building resilience, thereby stepping away from the image of ``helpless social media users and communities.''


\section{Discussion}
We introduced the \textit{folk models of misinformation on social media} as an effort to provide additional context to researchers studying misinformation rather than denigrating people who interact with it. There is not a ``correct'' model of misinformation that could serve as a comparison in the first place, therefore, we avoid giving preference to one model over another. The important aspect of our work is not how \textit{accurate} the model is, but how well it serves the needs of a social media user in dealing with misinformation \cite{Roy}. Misinformation differs, say from computer security \cite{Asgharpour} in that there is not yet a defined profile of an ``expert'' as facts could be from any domain, could be refuted in future, and there are competing, evolving contexts for interpreting an unbounded set of facts.  

This is perhaps where the illusory top-down approach for solving misinformation's wickedness lays. The self-appointed ``innoculators'' assume a \textit{de facto} misinformation expertise as they project their domain expertise to the problem of misinformation. Discrepancies are inherent---as the wicked problem definition predicts---because the problem of misinformation has collapsing and interwoven sets of (developing) facts from climate change, to politics, elections, public health and pandemics, and up to conflicts in foreign countries. No one has all the facts in the same time, nor the expertise to decide on the absolute ``truth-line,'' leaving users to not just untangle but also reject the misinformation. It seems ineffectual therefore to offer trainings, games, visual dashboards, and virtual reality rooms without having a clear picture on how the users---the ones in the front lines---conceptualize misinformation in the first place.  

The disconnectedness between the prescriptions for rejection and the symptoms of misinformation impedes an acceptance that misinformation, deceitfulness and ``fool me once'' attempts are integral part of our everyday life. A social media diet containing only pure, inherently factual, irrefutable facts seems flavorless and unlikely to be accepted by anyone involved. Misinformation might undoubtedly undermine the liberal democratic order, but---as our study reveals---it seems like the main ingredient in people's time on social media, and users do not necessarily want it removed altogether. Removing misinformation could perhaps be arranged, or at least aggressively enforced, by social media platforms, but that seems not to be their prerogative at all. The mainstream social media platforms need it for monetization, as our participants suspected, and the alternative ones for reprieve of the perceived free speech infringement \cite{gettr-paper}. 

This conundrum appears omitted in the top-down approach of misinformation containment and resilience. Structural factors are discussed, but hardly in any meaningful way of incentives individual social media users get when grappling with misinformation, beyond being driven by their personal values and beliefs. We brought the folk models to the front not just to show the emergence of ``natural immunity'' to misinformation, but to warn that it is in a very fragile state and runs the risk of being lost due to users' distrust in the these ``structures,'' discouragement in confronting misinformers, and a sense of being left to fight misinformation alone. And if this risk materializes, we are afraid, we have collectively been ``fooled for the n-teenth'' time.

\subsection{Ethics and Limitations}
The purpose of our project was not to generalize to a population; rather, to explore the phenomenon of personal dealing with misinformation in depth. To avoid misleading readers, we did not report definitive numbers of how many users possessed each folk model, nor how the folk models and the accompanying conceptualizations fair with participants' demographics. Instead, we describe the full range of folk models we observed, in a hope that the results can help to elevate the study of misinformation as a whole. One could say that there is a risk of oversimplification where the initial set of folk models of misinformation, expressed on the participants behalf, might not represent the entirety of folk models used to deal with misinformation. We, of course, acknowledge that there are certainly other ways and means that users employ and we welcome every work that brings them to the fore.   

Our research was limited by the size of our sample, and it was limited in its scope to American social media users. Each of these qualifications affected our understanding of how these folk-models for misinformation might be applied outside of this narrow scope. By asking users directly about how they interact with misinformation, we got a wide variety of fascinating insights from a broad range of different perspectives. We did not measure the efficacy of these folk models, or the variation between the results of different users' applications of folk models in a myriad of social media settings (users do have a preferred platform, but many of use several platforms interchangeably; also new social media platforms are regularly introduced). We are aware that these folk models represent the contextualizations and learned behavior informed by all forms of misinformation that currently exist on social media. Therefore, we are careful to avoid any predictive use of the folk models for future forms and problem domains of misinformation. 

Importantly, we do not know how our participants actually apply these folk models, as we did not observe them utilizing it in live settings, only how the participants reported that they apply them. In practice, many of these users might behave differently online, for example not checking facts, sources, commenting, reporting, or blocking/muting as as frequently as they said or in the order they mentioned. Participants, and users in general, could shift between models and vary the degree of analytical and heuristic assessment, so for example gut-checks, when applied by different individuals, might yield very different results from one user to the next. Also, we assumed the participants have undoubted trust in the fact checking process, however we acknowledge that the fact-checkers are not absolutely infallible in reality \cite{Brantzaeg, Walter}. 

There is a risk that this model could provide future research on misinformation with an easy citation that is over-reductive of the evolving experience of dealing with misinformation by users. We advise caution to this, as we see our work not as a direct alternative to the top-down approach, but rather as a synergistic line of scientific inquiry that produces a combined effect greater than the sum of their separate effects. We determined, however, that the lack of attempts by academics so far to research what individual users think about misinformation means that even a possibly incomplete or very limited analysis is better than what researchers currently have available to them to contextualize their future research on providing misinformation solutions. 

\subsection{Future Research}
Our future research will dive into each of these folk models to further explore how the overall misinformation folklore develops. Perhaps, as indicated by Wash \cite{Wash-folk}, there is a way to expose misinformation threats to individuals in a way that they understand (through use of their folk model(s)) and are motivated to take appropriate actions. Some potential future inquires should focus on extending the folk model of misinformation phenomenology 
and trace how individuals' resilience evolves/coalesces to create a sustained community-level resilience not just to external misinformation, but also from internal information-based threats. It would be particularly useful to study the circumstances in which the misinformation resilience defaults into information rejection and what costs and benefits the social media users incur for doing so. The last question deserves additional research to further understand each of the folk models and their ability to be altered, as the blanket rejection of (mis)information closes the door for dialogue. Because no dialog is a state of social media participation is what the ``foolers'' might want the most. 


\section{Conclusion}  \label{section:7}
What misinformation is, or represents, undoubtedly evolved from the mid 2010's until today. The fake news fabricated by foreign actors are not anymore the sole descriptor of misinformation as misinformation became a state of mind. A testimony to this evolution are the folk models of misinformation identified in this study, as well as the what users employ to ensure they preserve their well-being despite being regularly exposed to misinformation on social media. The folk models of misinformation reveal a natural tendency and motivation by social media users to emulate an ``expert'' role in dealing in misinformation, effectively creating conditions for a resilience through ``natural immunity'' to take effect.

\bibliographystyle{ACM-Reference-Format}
\bibliography{\jobname}

\appendix

\section{Survey Questions}
\subsection{Exposure and Preconceptions}
\begin{enumerate}[start=0, label=\arabic*.]
\itemsep 1em
    \item Misinformation on social media is an umbrella term that includes all false or inaccurate information that is spread on social network platforms, such as: disinformation, fake news, rumors, conspiracy theories, hoaxes, trolling, urban legends, and spam \cite{Wu}. 
    
    \item Taking the above definition in consideration, could you please specify all the platforms where you have encountered misinformation and, if possible, provide some examples \textbf{[Twitter, Facebook, Reddit, Gab, Gettr, Parler, Rumbler, Truth Social]}
    
    \item What was your initial response to some of the misinformation examples you have provided? Please elaborate. \textbf{[Open Ended]}
    
    \item Where does misinformation on social media come from, in your opinion?
    \textbf{[Open Ended]}

    \item What kind of function does the misinformation on social media serve, in your opinion?
    \textbf{[Open Ended]}
    
    \item Who benefits from misinformation on social media, in your opinion?
    \textbf{[Open Ended]}

    
\end{enumerate}

\subsection{Engagement Strategies}
\begin{enumerate}[resume, label=\arabic*.]
\itemsep 1em
    \item How do you suspect/know a certain social media post is misinformation? Please elaborate. \textbf{[Open Ended]}
    
    \item What is your strategy for dealing with misinformation posts on social media? Please elaborate. \textbf{[Open Ended]}
    
    \item Are there occasions where you have, or are inclined to, comment/reply to a misinformation post? If so, what did you or would you say in your comment/reply? Please elaborate \textbf{[Open Ended]}
   
    \item Are there occasions where you have, or are inclined to, use any engagement features (e.g. like, retweet/repost, share, follow) when encountering a misinformation post? If so, in what circumstances? Please elaborate \textbf{[Open Ended]}
    
    \item Are there occasions where you have, or are inclined to, use any action features (e.g. block, mute, report, unfollow) when encountering a misinformation post? If so, in what circumstances? Please elaborate \textbf{[Open Ended]}
    
    \item Are there occasions where you have or are inclined to talk about a particular misinformation post outside social media? If so, in what circumstances? Please elaborate \textbf{[Open Ended]}
    
    \item Are there occasions where you have or are inclined to engage with a misinformation post using counter-argumentation? If so, in what circumstances? Please elaborate \textbf{[Open Ended]}
    
    \item Are there occasions where you have or are inclined to engage with a misinformation post using humor, sarcasm, mocking, or taunting? If so, in what circumstances? Please elaborate \textbf{[Open Ended]}

\end{enumerate}

\subsection{Practice: Randomized Exposure}
\begin{enumerate}[resume, label=\arabic*.]
\itemsep 1em
    \item Consider this post \textbf{[Downright COVID-19/political/Ukraine War conspiracy]}. In your opinion is this post misinformation? How would you respond to it, assuming this post showed in your social media feed? Please elaborate on both accounts \textbf{[Open Ended]}
    
    Putin claims he got the idea to invade Ukraine after watching an episode of The Americans and thinking back to the good old days of being a spy.
    
    \item Consider this post \textbf{[Misleading COVID-19/political/Ukraine War statistics post]}. In your opinion is this post misinformation? How would you respond to it, assuming this post showed in your social media feed? Please elaborate on both accounts \textbf{[Open Ended]}
    
    Elon Musk is buying up America's brands promising to make them "better". Twitter will be back to allowing all speech to be free and Coca-Cola will return to it's original formula with cocaine.
    
     \item Consider this post \textbf{[Obscure but truthful COVID-19/political /Ukraine War post]}. In your opinion is this post misinformation? How would you respond to it, assuming this post showed in your social media feed? Please elaborate on both accounts \textbf{[Open Ended]}
    
    The Helms Amendment has been used to restrict/prevent abortion treatments for women violated during the war in Ukraine.  https://tinyurl.com/y67cbcff
    
\end{enumerate}

\subsection{Wrap Up and Follow Up}
\begin{enumerate}[resume, label=\arabic*.]
\itemsep 1em
    \item We would like to hear anything that you want to say about how you, or the people you know, handle misinformation on social media \textbf{[Open Ended]}
    
    \item We plan to follow-up with anyone that is down for a zoom interview on personal takes when it comes to dealing with misinformation. If you like to do so, please give us the best way to contact you. \textbf{Contact(email, phone, social media): [Open Ended]}

\end{enumerate}

\end{document}